\documentclass[conference, letterpaper]{IEEEtran}
\usepackage{cite}
\usepackage{amsmath,amssymb,amsfonts}
\usepackage{algorithmic}
\usepackage{graphicx}
\usepackage{textcomp}
\usepackage{xcolor}
\def\BibTeX{{\rm B\kern-.05em{\sc i\kern-.025em b}\kern-.08em
    T\kern-.1667em\lower.7ex\hbox{E}\kern-.125emX}}
\usepackage[left=1.62cm,right=1.62cm,top=1.9cm]{geometry}

\begin{document}

\title{Flight Testing an Optionally Piloted Aircraft: a Case Study on Trust Dynamics in Human-Autonomy Teaming}

\author{
\IEEEauthorblockN{1\textsuperscript{st} Jeremy C.-H. Wang}
\IEEEauthorblockA{\textit{Ribbit}\\
Toronto, Canada\\
jeremy.wang@flyribbit.com}

\and

\IEEEauthorblockN{2\textsuperscript{nd} Ming Hou}
\IEEEauthorblockA{\textit{Defence R\&D Canada}\\
Toronto, Canada\\
ming.hou@forces.gc.ca}

\and

\IEEEauthorblockN{3\textsuperscript{rd} David Dunwoody}
\IEEEauthorblockA{\textit{Royal Canadian Air Force}\\
Ottawa, Canada\\
david.dunwoody@forces.gc.ca}

\and[\hfill\mbox{}\par\mbox{}\hfill]

\IEEEauthorblockN{4\textsuperscript{th} Marko Ilievski}
\IEEEauthorblockA{\textit{Ribbit}\\
Toronto, Canada\\
marko.ilievski@flyribbit.com}

\and

\IEEEauthorblockN{5\textsuperscript{th} Justin Tomasi}
\IEEEauthorblockA{\textit{Ribbit}\\
Toronto, Canada\\
justin.tomasi@flyribbit.com}

\and

\IEEEauthorblockN{6\textsuperscript{th} Edward Chao}
\IEEEauthorblockA{\textit{Ribbit}\\
Toronto, Canada\\
edward.chao@flyribbit.com}

\and

\IEEEauthorblockN{7\textsuperscript{th} Carl Pigeon}
\IEEEauthorblockA{\textit{Ribbit}\\
Toronto, Canada\\
carl.pigeon@flyribbit.com}
}

\maketitle

\begin{abstract}
This paper examines how trust is formed, maintained, or diminished over time in the context of human-autonomy teaming with an optionally piloted aircraft. Whereas traditional factor-based trust models offer a static representation of human confidence in technology, here we discuss how variations in the underlying factors lead to variations in trust, trust thresholds, and human behaviours. Over 200 hours of flight test data collected over a multi-year test campaign from 2021 to 2023 were reviewed. The dispositional-situational-learned, process-performance-purpose, and IMPACTS homeostasis trust models are applied to illuminate trust trends during nominal autonomous flight operations. The results offer promising directions for future studies on trust dynamics and design-for-trust in human-autonomy teaming.
\end{abstract}


\section{Introduction}

How do humans develop the confidence to use emerging technologies---that is, how do we learn to \emph{trust} that which is new? Studies show that a myriad of factors produce statistically significant empirical effects on the degree to which humans are willing to depend on technology, such as the expertise and demographics of the trustor, the reliability and anthropomorphism of the trustee, and the difficulty and risk of the task at hand \cite{HANCOCK_ET_AL_2011, HOU_ET_AL_2014, HANCOCK_ET_AL_2020, HOU_ET_AL_2024}. These factors are often grouped into categories for the purpose of analysis, yielding various \emph{trust models} such as the performance-process-purpose (3P) model \cite{LEE_SEE_2004} and dispositional-situational-learned (DSL) model \cite{HOF_BASHIR_2015}.

Safety is arguably the most important factor that affects human trust in emerging aviation technologies, especially in the case of aircraft automation. Regulators set safety requirements across the entire aircraft life cycle from design through to production, operation, and maintenance \cite{FAR_2024, CAR_2024, EASA_2018}. When regulators deem these requirements satisfied, they issue certificates, permits, or other formal approvals that help instil public confidence \cite{SIX_VERHOEST_2017}. Garmin Autoland, which can automatically land an aircraft if it determines the pilot is unable to fly, offers a contemporary example of celebrated automation \cite{GARMIN_2021}. Conversely, major incidents such as the Boeing 737 MAX \cite{737MAX_2020} crashes can significantly damage confidence in specific technologies, not to mention the organizations charged with their development and regulation. The highly dynamic and contextual nature of trust also manifests in other safety-critical contexts, for instance, the Cruise robo-taxi incident \cite{KOOPMAN_2024}. Trust can thus be earned, maintained, or lost in response to underlying factors.

Although trust factors have been established in the literature, the continuously varying nature of trust over time is not well-studied. In an era of increasingly complex and automated aircraft \cite{ALEXIEV_ET_AL_2024, JANSEN_ET_AL_2024}, human factors engineering would benefit from research that investigates the highly dynamic relationship between humans and their willingness to depend on emerging technologies. Indeed, the study of trust variation and evolution has been proposed as one of the key enablers for effective Human-Autonomy Teaming (HAT) \cite{HOU_ET_AL_2021, PUSCAS_2022, HOU_2023}. Hou \emph{et al.}'s recent IMPACTS Homeostasis (IMPACTS-H) trust model offers a promising conceptual framework that considers trust as a changing quantity whose distance from some threshold influences how human operators interact with intelligent systems \cite{HOU_ET_AL_2021, HOU_ET_AL_2024}.

This working paper seeks to narrow the knowledge gap in dynamic trust by studying human behaviours observed during flight testing of an optionally piloted aircraft (OPA). OPAs are aircraft that may be flown in an uncrewed or crewed fashion, thus offering the flexibility to assess autonomous capabilities with onboard or remote human supervision. The 3P and DSL and IMPACTS-H trust models are used to analyze flight test crew behaviours, shedding light on the variable nature of human confidence in HAT.

\section{Experimental Design}

Modern autopilots are no longer "dumb and dutiful", nor are they limited to the simplistic designs used in small recreational or commercial drones \cite{WANG_2022_IEEE}. Instead, contemporary autopilots act evermore as a partner that complements their human operator counterparts. Historical data of numerous flight tests were analyzed to reveal the time evolution of trust by personnel involved in this emerging technology. The original purpose of these flight tests was the development, performance evaluation, and demonstration of autonomous flight capabilities---for the purpose of this paper, the same flight test data was re-analyzed in the context of trust in HAT.

\subsection{Equipment}

\begin{figure}[h]
\centerline{\includegraphics[scale=0.13]{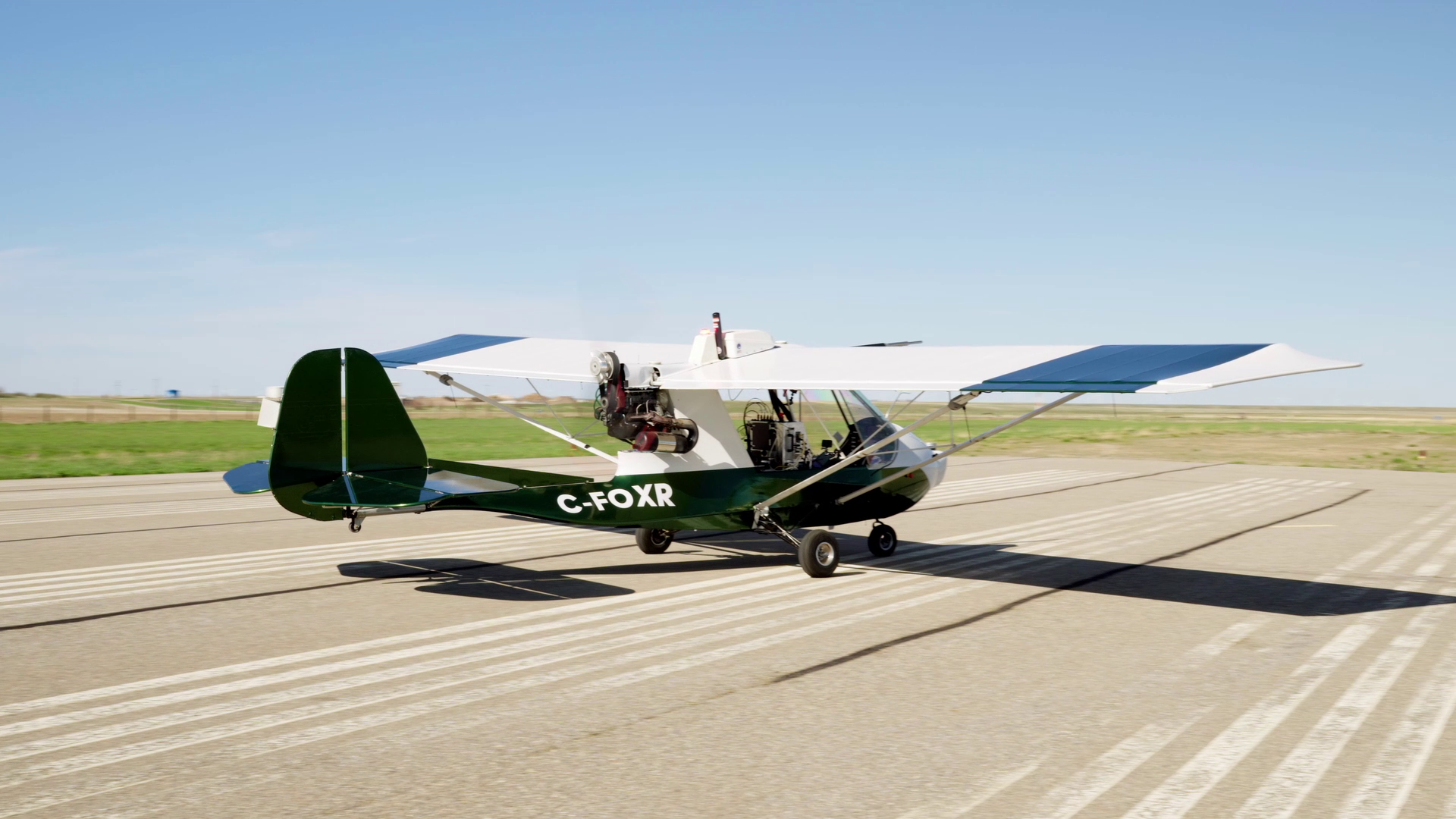}}
\caption{The OPA used in this study, shown during remotely supervised autonomous flight tests near Canadian Forces Base Cold Lake.}
\label{fig:opa}
\end{figure}

\begin{figure}[h]
\centerline{\includegraphics[scale=0.13]{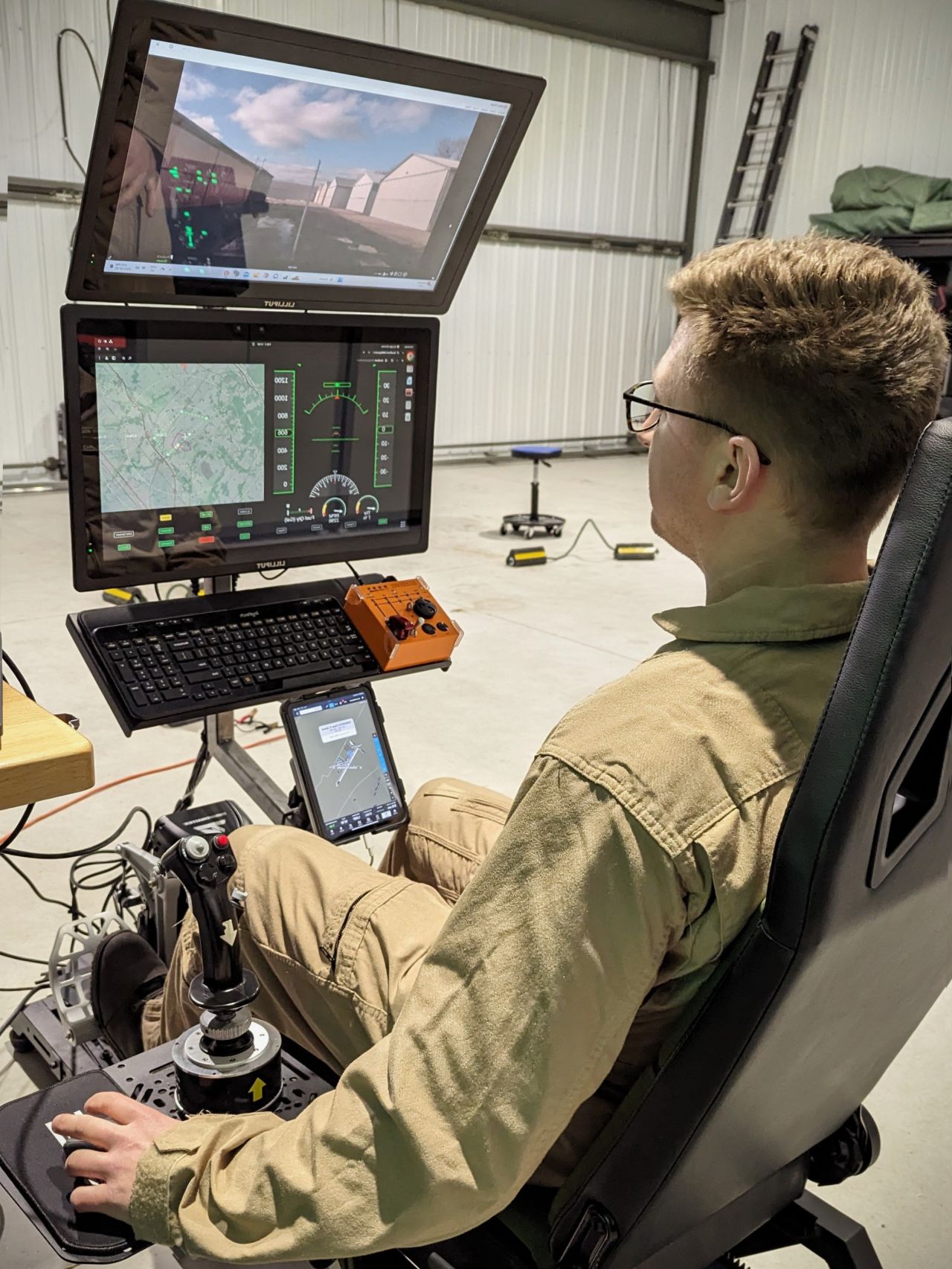}}
\caption{The GCS used by remote pilots and flight test engineers to monitor and interact with the OPA from a ground-based setting.}
\label{fig:gcs}
\end{figure}

The Ribbit autonomy stack was integrated aboard a two-seat Quad City Challenger 2 airplane (Fig. \ref{fig:opa}), resulting in an OPA capable of automated gate-to-gate flight operations from runway-based airports. This technology is being developed for the purpose of improving access to transportation in hard-to-reach areas, \emph{e.g.} for civil air freight \cite{WANG_ET_AL_2025}, food resupply \cite{GUAN_WANG_2024}, as well as force projection and sustainment. Aircraft state was measured and recorded through several onboard sensors, such as an air data boom,  inertial measurement unit, and cockpit camera. A Ground Control Station (GCS) also permitted remote pilots and/or flight test engineers to monitor aircraft state live from afar (Fig. \ref{fig:gcs}). All designs, modifications, and flight tests were conducted within the framework of the Canadian Aviation Regulations \cite{CAR_2024}. When operated with an onboard safety pilot, the pilot was capable of monitoring various aircraft state variables as well as engaging or disengaging the autonomous autopilot to restore stick-and-rudder control. When operated in a remotely piloted fashion, the pilot was able to assume manual control via fly-by-wire inceptors on the GCS.

\subsection{Test Procedures}

Flight tests were conducted over a three-year period from 2021 to 2023, accumulating some 200 hours of aircraft-in-motion data spanning autonomous taxi, takeoff, enroute, and landing test sequences. Significant variation in environmental conditions was included by design, including summer and winter operations, and flight tests up to the baseline aircraft's maximum demonstration crosswind. For safety and efficiency, all tests were planned in advance and flown in simulation prior to execution on the physical aircraft. The purpose of flight tests were generally for benchmarking (\emph{e.g.} establishing a human performance baseline), system identification (\emph{i.e.} generating a physics model of aircraft dynamics and stability), system tuning (\emph{e.g.} refining controller and navigation parameters to achieve desired performance), verification of design requirements, and demonstration to third-party observers.

\subsection{Data Generation, Analysis, \& Interpretation}

The trust levels and thresholds of three (3) different test engineers and two (2) different test pilots are considered in this study. These participants were asked to review historical time-series data consisting of time-synchronized cockpit videos, flight displays, and flight test engineering graphs of key aircraft state variables (\emph{i.e.} setpoint and process values for outer loop and inner loop control variables) from the aforementioned test campaign. It is important to note that while these participants are the same individuals who conducted the original flight tests, their self-reported trust levels and thresholds were not contemporaneously recorded. Instead, participants subjectively interpreted and reported their trust levels and thresholds after the fact based on their review and interpretation of the data (\emph{e.g.} number and timing of autopilot disengagements, precautionary body language, error between setpoint and process variables related to trajectory following). The resulting trust curves for sample test sequences are discussed in this paper, where trust is normalized on a scale of 0 (none whatsoever) to 1 (absolute trust).

Three trust models were then used to analyze trust dynamics: 3P \cite{LEE_SEE_2004}, DSL \cite{HOF_BASHIR_2015}, and IMPACTS-H \cite{HOU_ET_AL_2024}. 3P factors are helpful when changes in trust are driven by perceived differences in human versus machine performance whereas DSL factors are helpful when distinguishing preceding (dispositional), live (situational), and cumulative (learned) contributors to trust. Meanwhile, IMPACTS-H analysis provides a framework to visualize and analyze trust values, thresholds, comparisons, and behaviours.

Additionally, the IMPACTS elements of IMPACTS-H (intention, measurability, performance, adaptivity, communication, transparency, and security) were considered during the development of Ribbit's autonomy stack, thus permitting the complex latent trust dynamics of aviation HAT to be observed. These different analytical lenses are used through the discussion of results to elucidate homeostatic properties observed in the context of HAT with an OPA. 

\subsection{Limitations}

Since tests and data collection were not originally designed with trust research in mind, trust perceptions and interpretations were inferred \emph{a posteriori}. Follow-on studies designed specifically for trust evaluation are planned for the future. Additionally, due to the proprietary nature of the technology, only certain details can be disclosed in this contribution.

\section{Results \& Discussion}

A nominal gate-to-gate flight can be divided according to major phases of flight and the transitions between them: taxi, takeoff, enroute, and landing. Furthermore, special state transitions exist such as rejected takeoffs, go-arounds, and precautionary landings which may be triggered in response to abnormal operating conditions. The distinct operational requirements and safety risks therein lend themselves naturally to discussing trust dynamics by each phase.

\subsection{Taxi}

\begin{figure}[h]
\centerline{\includegraphics[scale=0.62]{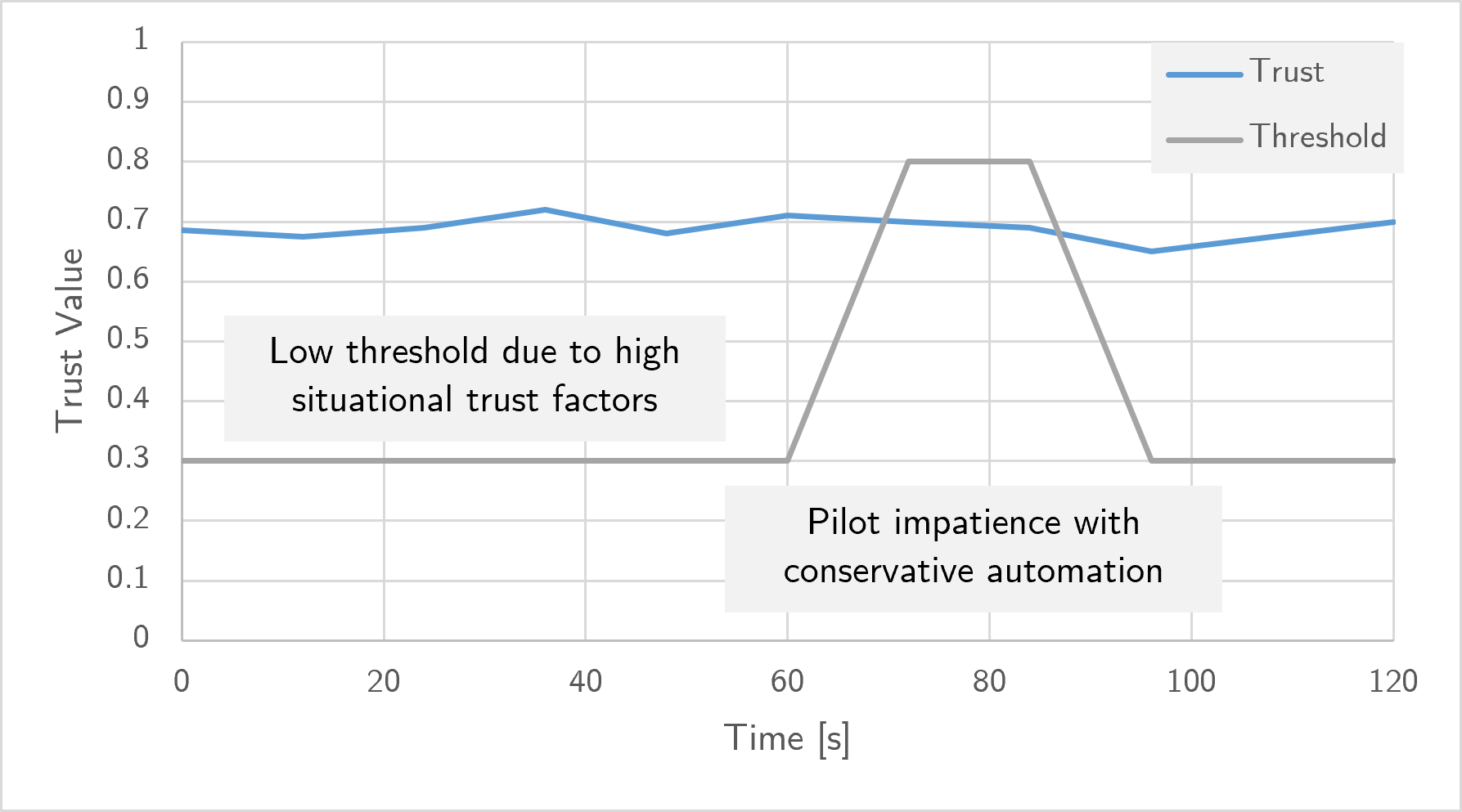}}
\caption{Test pilot trust dynamics during a sample taxi sequence, highlighting a relatively stable trust curve and a visualization of test pilot impatience impacting the trust threshold. When the threshold exceeds the trust, human intervention occurs.}
\label{fig:taxi}
\end{figure}

Ground taxi operations presented favourable situational trust factors. Pilots and engineers perceived low risk, low workload, and limited system complexity thanks to the low vehicle speeds, simple control behaviours, slow aircraft physical responses, and fixed taxiways and traffic patterns. Even with self-reported dispositional differences in risk tolerance between crew members, the trust threshold required to proceed with autonomous operation was perceived to be very low. Still, the test pilot sometimes preferred to taxi manually when the primary purpose of the test was unrelated to the taxi phase---this is because the autonomy stack was generally configured with gentler acceleration and braking compared to the faster, more anthropomorphic, driving style of human pilots. Anthropomorphism is thus a trust factor that can cause pilot and engineer trust evolution to behave differently: flight test engineers see the system performing as it should, whereas the test pilot may find this nominal performance to be unsatisfactory. An example of trust during taxi based on these factors is shown in Fig. \ref{fig:taxi}.

\begin{figure}[h]
\centerline{\includegraphics[scale=0.62]{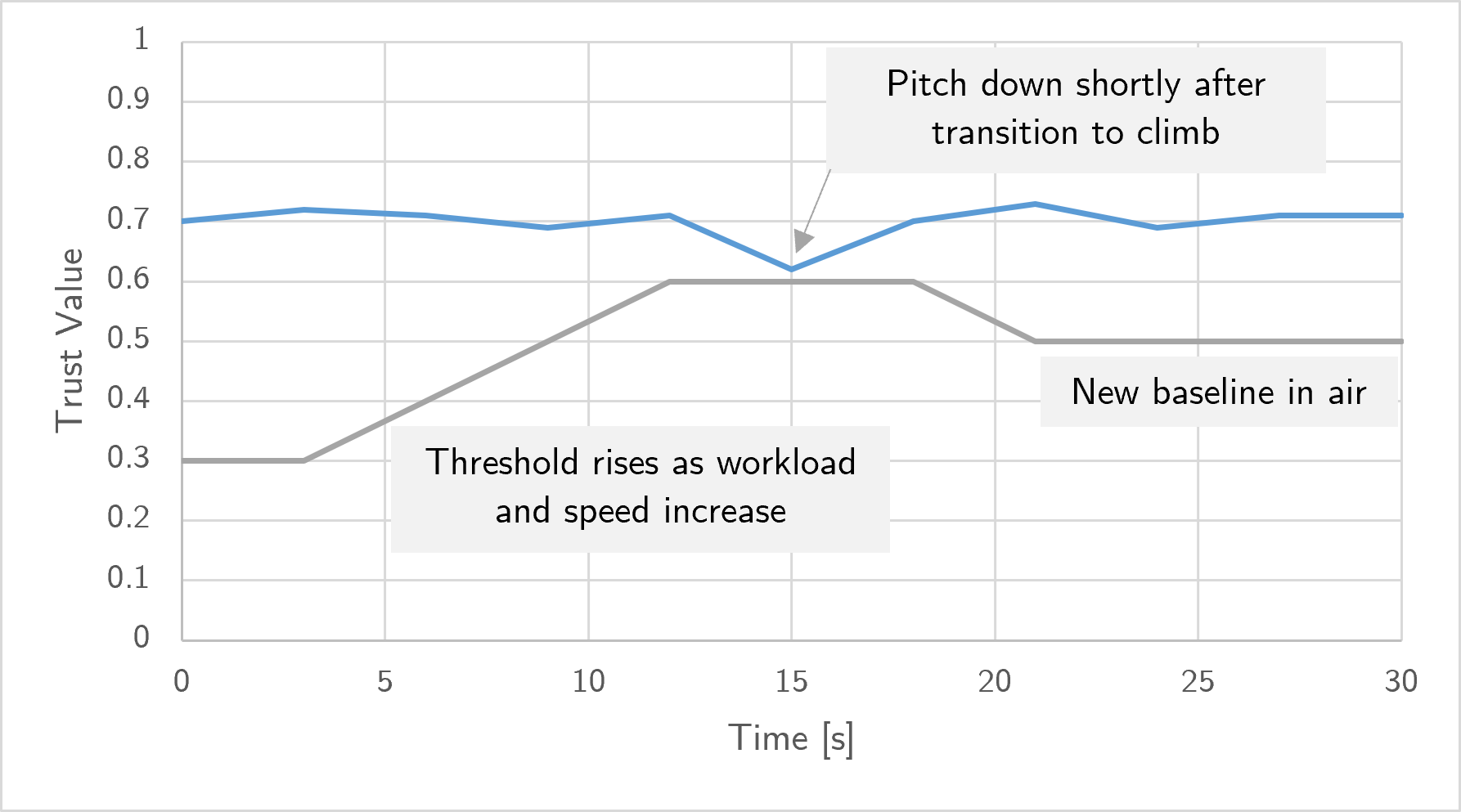}}
\caption{Crew member trust dynamics during a sample takeoff sequence. The threshold steadily rises during the takeoff roll owing to increased perceived risk and workload, before eventually re-establishing along a new baseline for flight.}
\label{fig:takeoff}
\end{figure}

\subsection{Takeoff}

\begin{figure}[h]
\centerline{\includegraphics[scale=0.55]{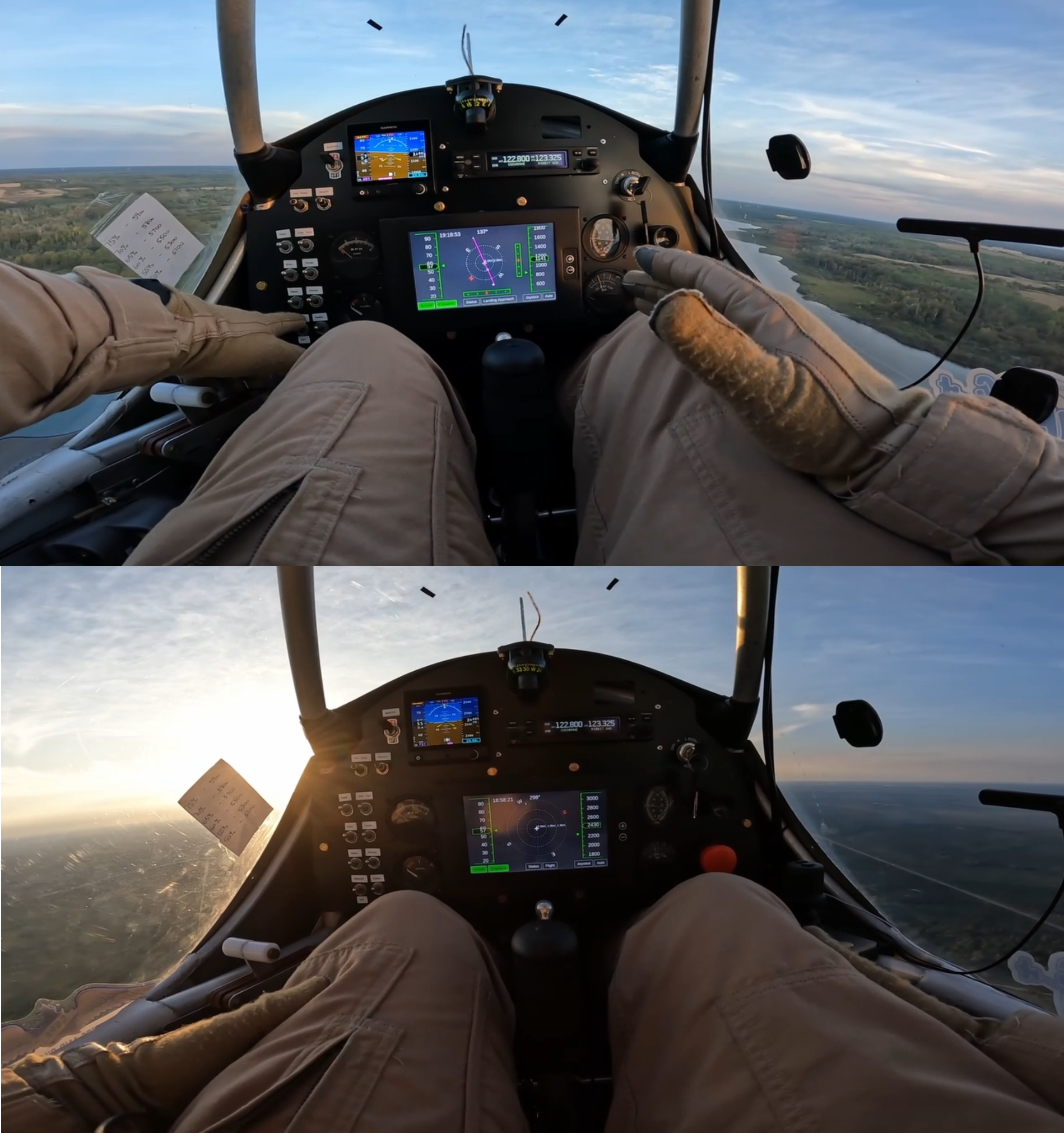}}
\caption{Comparison of pilot hand positions during periods of increased perceived risk (top, such as during takeoff or landing) versus decreased perceived risk (bottom, such as during taxi or enroute).}
\label{fig:hands}
\end{figure}

During takeoff, test pilots and engineers perceived increased risk due to the aircraft's rising speed and the onboard sensation of higher accelerations and vibrations. Task difficulty, and thus workload, also increased as crew members carefully monitored speeds, centreline tracking, engine variables, remaining runway distance, radio transmissions, and potential runway obstructions or debris that could warrant a rejected takeoff. The presence of any steady or gusting winds added to external variability, thus raising the trust threshold further. Perhaps most critically, the control laws used during the takeoff roll were expected to smoothly and safely transition to those used during a full-power climb. In the moments after lift-off, any sudden downward pitching moment---even if objectively safe and stable---raised pilot stress levels and incurred distrust. The elevated trust threshold throughout the takeoff roll and initial climb was reflected in the hand positions of the test pilot: hovering over the disengage switch and ready to assume manual control (Fig. \ref{fig:hands}). Once established in the climb and at a sufficient altitude to permit safe recovery from a stall or engine failure, the trust threshold decreases until a new baseline for in-air maneuvers is established. These variations are visualized in Fig. \ref{fig:takeoff}.

\subsection{Enroute}

The enroute phase of flight generally lasted the longest of all flight phases, permitting the examination of long-term trends whilst trust factors remained relatively stable. The autonomy stack was normally expected to hold steady-level flight or conduct turns, climbs, and descents to manage the track, speed, and altitude of the aircraft. The autonomous system always outperformed the test pilot with respect to ride comfort and precise flight path management, which when observed consistently, led to a slow and gradual increase in human trust before reaching yet another, higher baseline. It is likely that learned trust factors, namely prior exposure to traditional autopilots, enhanced trust in the autonomy stack even prior to first contact---this previous exposure may have also resulted in increased performance expectations, and hence an increased trust threshold as well. 

A notable exception to the pro-autonomy trust behaviour was observed during system identification flights. During such flight tests, it was necessary to conduct frequency sweeps along different degrees of freedom in order to extract aircraft stability characteristics. Test pilots were more willing to undergo greater amplitudes and frequencies when they were flying versus the autonomy stack. This preference defied the factual knowledge that the autonomy system was capable of finer control, which with proper planning, would still avoid exceeding the flight envelope. Nevertheless, one could argue the subjective sense of control and physiological stress of rollercoaster-like motions led to decreased trust in the autonomy stack. This trust behaviour has been documented in texts on system identification flight test planning \cite{TISCHLER_REMPLE_2006}.

\begin{figure}[h]
\centerline{\includegraphics[scale=0.62]{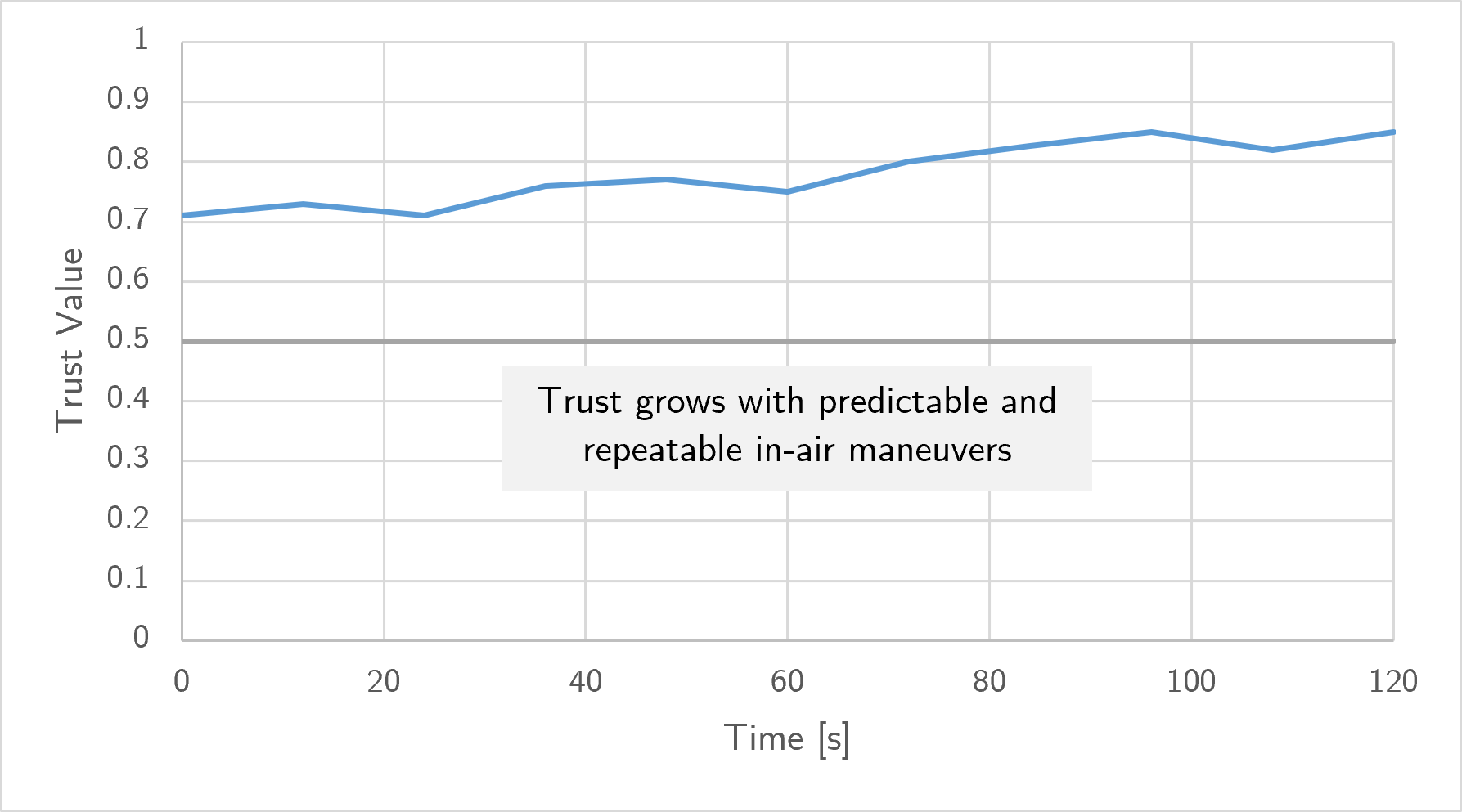}}
\caption{Crew member trust inflation during a sample enroute test sequence, during which steady-level cruise, turns, climbs, and descents occur. Trust grows as the pilot becomes increasingly comfortable owing to sustained desirable behaviours, prior learned familiarity with traditional autopilots, and superior performance of the autonomy stack compared to human benchmarks.}
\label{fig:enroute}
\end{figure}

\subsection{Landing}

\begin{figure}[h]
\centerline{\includegraphics[scale=0.62]{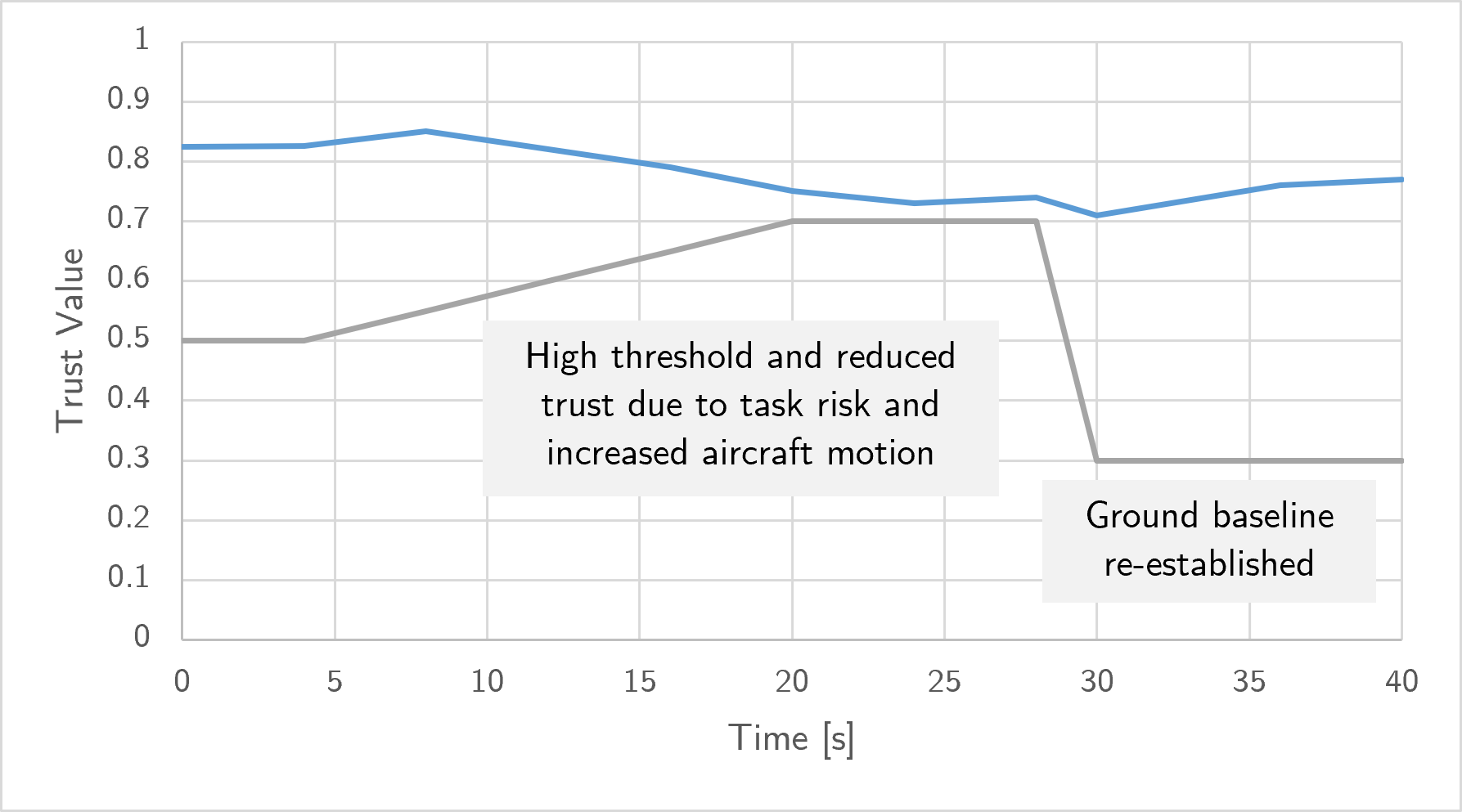}}
\caption{Crew member trust dynamics during a sample landing test sequence, showing high trust threshold due to task risk, difficulty, and sensation of aircraft motion---followed quickly by re-establishing along ground baselines once fully on the ground and off the runway.}
\label{fig:landing}
\end{figure}

Landing may be decomposed further into individual subphases, such as the approach, flare, and rollout. In this sense, the situational and process trust factors that applied to takeoff also apply to landing---except that the aircraft is now attempting to shed energy, which is generally more challenging than adding to total energy via engine power. The landing sequence also tended to involve more aircraft motion, as the aircraft was required to pursue a precise yet stable approach toward the ground. During flare, engine power was removed to permit the aircraft to touch down gently while simultaneously degrading a degree of freedom. Thus, the probability of a hard touchdown or ballooning was largely mitigated in advance via a stabilized approach that set the aircraft up for a successful touchdown. Together, the process, purpose, risk, sensation, complexity, and controllability of landing all resulted in more stringent performance requirements. Consequently, the trust threshold rose more quickly and reached the highest values during the landing phase (Fig. \ref{fig:landing}). Once on the ground and off the runway, the trust and trust thresholds re-baselined toward taxi levels.

\section{Conclusion}

The study of HAT in OPA operations presents a fertile setting for continued research on trust dynamics in safety-critical contexts. By exploring trust factors according to different phases of flight, we showed how trust and trust thresholds vary based on changes in factors captured in the DSL, 3P, and IMPACTS-H trust models. The homeostatic nature of trust was explored in terms of changes to trustor, trustee, and situational factors, as well as routine re-baselining under quasi-static conditions. It is evident that while factors-based approaches aid the interpretation of observed trust behaviours, they remain insufficient for generating detailed, quantitative predictions of trust measures. Future studies on HAT in aviation should focus on defining quantitative trust measures, developing dynamic models, and testing these measures and models under controlled circumstances such as specific and repeatable flight maneuvers.

\bibliographystyle{ieeetr}
\bibliography{lib-master.bib}
\vspace{12pt}

\end{document}